\documentclass[preprint]{aastex}
\usepackage{natbib}
\begin{document}

\title{Monitoring the Sky with the Prototype All-Sky Imager on the LWA1}
\author{K.S. Obenberger \\ \affil{Department of Physics and Astronomy, University of New Mexico, Albuquerque, NM 87131} G.B. Taylor\\  \affil{Department of Physics and Astronomy, University of New Mexico, Albuquerque, NM 87131}  J.M. Hartman\\ \affil{Google, Venice, CA 90291} T.E. Clarke\\  \affil{US Naval Research Laboratory, Code 7213, Washington, DC 20375} J. Dowell\\  \affil{Department of Physics and Astronomy, University of New Mexico, Albuquerque, NM 87131} A. Dubois\\ \affil{Los Alamos National Laboratory, Los Alamos, NM 87545} D. Dubois\\ \affil{Los Alamos National Laboratory, Los Alamos, NM 87545} P.A. Henning\\  \affil{Department of Physics and Astronomy, University of New Mexico, Albuquerque NM, 87131} J. Lazio\\ \affil{Jet Propulsion Laboratory, California Institute of Technology, Pasadena, CA 91109} S. Michalak, \\ \affil{Los Alamos National Laboratory, Los Alamos, NM 87545} F.K. Schinzel\\  \affil{Department of Physics and Astronomy, University of New Mexico, Albuquerque, NM 87131}   }


\begin{abstract} We present a description of the Prototype All-Sky Imager (PASI), a backend correlator and imager of the first station of the Long Wavelength Array (LWA1). PASI cross-correlates a live stream of 260 dual-polarization dipole antennas of the LWA1, creates all-sky images, and uploads them to the LWA-TV website in near real-time. PASI has recorded over 13,000 hours of all-sky images at frequencies between 10 and 88 MHz creating opportunities for new research and discoveries. We also report rate density and pulse energy density limits on transients at 38, 52, and 74 MHz, for pulse widths of 5 s. We limit transients at those frequencies with pulse energy densities of  $>2.7\times 10^{-23}$, $>1.1\times 10^{-23}$, and $>2.8\times 10^{-23}$ J m$^{-2}$ Hz$^{-1}$ to have rate densities $<1.2\times10^{-4}$, $<5.6\times10^{-4}$, and $<7.2\times10^{-4}$ yr$^{-1}$ deg$^{-2}$.

\end{abstract}

\section{Introduction}
A large variety of astronomical sources have been theorized to produce transient, non-thermal radio emission on timescales of milliseconds to minutes. Among these sources are Gamma-Ray Bursts (GRBs; \citealt{Usov00,Sagiv02}), neutron star mergers \citep{Hansen01}, primordial black holes \citep{Rees77,Blandford77,Kavic08}, flaring stars \citep{Loeb13}, and magnetospheric emissions from hot Jupiters \citep{Stevens05}. Many of these sources are thought to emit strongly below 100 MHz with steep spectral indices. Moreover newly discovered fast radio bursts (FRBs) near 1.4 GHz provide an example of known transient emission, which displays a steep spectrum providing the potential for expanded studies at low frequencies \citep{Lorimer07,Keane12,Thornton13,Spitler14}.  

Recently a search at 142 MHz with the Low Frequency Array (LOFAR) limited transients on timescales of 0.7 ms \citep{Coenen14}.  Also a search with the Murchison Widefield Array (MWA) at 154 MHz has limited transients of timescales of 30 minutes \citep{Bell14}. However, to date there have been very few blind searches for transients below 100 MHz \citep{Kardashev77,Lazio10,Cutchin11}, and none of these studies have resulted in a detection. 

The first station of the Long Wavelength Array (LWA1; Ellingson 2013; Taylor 2012) is a radio telescope which operates between 10 and 88 MHz. The telescope consists of 256 dual-polarization dipole antennas \citep{Hicks10} distributed within a 100 $\times$ 110 m ellipse, with 5 additional outlier antennas with distances between 200 to 500 m from the center of the main array. The array is collocated with the Karl Jansky Very Large Array (VLA) at a latitude of 34.070$^{\circ}$ N and a longitude of 107.628$^{\circ}$ W.


The beam pattern of each LWA1 antenna is sensitive to nearly the entire sky and, therefore, delay tracking for specific sources is not necessary. Rather since all-sky images are the desired result, the array can simply be phased to zenith. This fact makes interferometry with a dipole-based array simpler than a typical array using dishes.

Several observing programs have recorded a small amount of raw LWA1 data and correlated it for different scientific applications, including ionospheric and meteor studies \citep{Helmboldt13,Helmboldt14}. However, since the raw data rate is so large (357 GB/hr), recording it to disk for later correlation is not feasible for large scale, blind transient searches, which are one of the main science drivers of the LWA1. Instead realtime correlation and imaging is the preferred method of observing, because it enables a large reduction in the data rate.

The Prototype All-Sky Imager (PASI) is a backend to the LWA1 that correlates and images the live stream of all LWA1 antennas and uploads the images to the LWA-TV\footnote{http://www.phys.unm.edu/$\sim$lwa/lwatv.html} website in near real-time. The all-sky images from PASI are stored in an archive where about 13,000 hours of data have been accumulated. Previous and ongoing projects have used this data to discover non-thermal radio emission from large meteors (fireballs; \citealt{Obenberger142}) and place limits on both prompt low frequency radiation from GRBs and general transients \citep{Obenberger141}. 

This paper describes the PASI system and provides improved limits on transients at 38, 52, and 74 MHz. The paper is organized as follows: \S2 describes the PASI computing hardware. \S3 describes the correlation and imaging software. \S4 describes the data output from PASI and its calibration and analysis. \S 5 discusses the limits PASI provides on astronomical transients. \S 6 discusses the outreach and education benefits of PASI. Finally, in \S 7 we present our conclusions. 

\section{Hardware}

The PASI computing cluster consists of 4 nodes. All nodes contain two quad-core Nehalem 2.93 GHz processors and 12 GB of RAM, except for node 4 which has 14 GB of RAM. The nodes are connected using an Infiniband switch. Processing tasks are divided between the nodes. Node 1 reads the TBN stream and handles the polyphase FFT, Node 2 runs the correlator, and Node 4 runs the imager, leaving Node 3 for user post-processing and as a spare node in case one of the others irreparably fails. During operation the 15 minute averaged processing loads\footnote{A load number of 0 represents an idle computer, any process using or waiting for CPU increments the load number by 1.} on each of the nodes are as follows: Node 1 $\sim$ 4.7, Node 2 $\sim$ 0.2, and Node 4 $\sim$ 1.8. Since each node has two quad core processors with 2 threads per core, none of the nodes are overloaded.

The cluster has an internal storage capacity of 7.5 TB, with eSATA ports for connecting external storage as needed. Currently PASI is connected to 17 TB of external storage, 15 TB of which are dedicated to a ring buffer for visibilities. The other 2 TB are used as a swappable image archive. Output data from PASI can be accessed on the external hard drives and be copied over fiber to other machines. PASI can also be controlled remotely and used for post-processing without having to copy the data to other machines. 

PASI is located within the electronics shelter of the LWA1. Figure 1 shows a flow chart that describes how PASI interfaces with the LWA1 system. The signals received from each antenna are impedance matched and pre-amplified by the front end electronics (FEE) at each stand. The generated signals are then passed on to the analog receiver boards (ARX) inside the electronics shelter where they are further amplified and filtered. After that, the digital processor (DP) converts the analog signals to digital complex voltages for further processing.

\begin{figure}
	\centering
	\includegraphics[width = 7in]{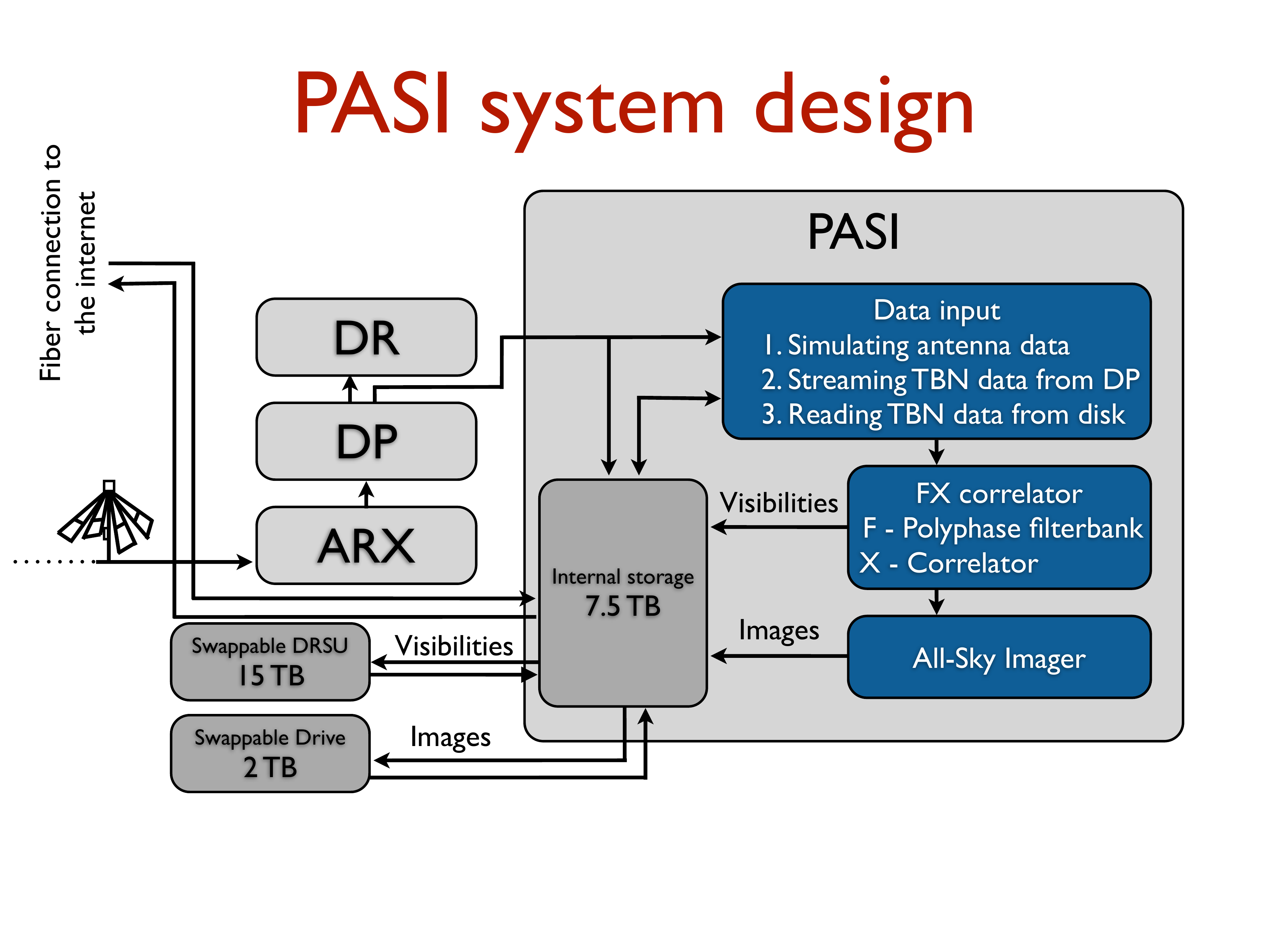}
	\caption{A block diagram showing the PASI components and other relevant LWA systems. Dashed lines indicate command and status information pathways and are carried by standard Ethernet connections. Solid lines indicate signal or high-bandwidth data pathways and are carried by 10 GbE connections outside of PASI and Infiniband connections or shared memory within it. Grey boxes represent LWA1 hardware systems, and  blue boxes represent components of the PasiFx software. A description of the acronyms are as follows: DR (Data Recorder), DP (Digital Processor), ARX (Analog Receiver), DRSU (Data Recorder Storage Unit), and TBN (Transient Buffer Narrow band) }
\end{figure}

DP can operate in two modes: digital beam forming and transient buffer \citep{Ellingson13}. The transient buffer provides the linearly-polarized voltage time series from all antenna stands, and has two submodes: wideband and narrowband. Wideband (TBW) allows for the entire 78 MHz bandwidth output from the ARX to be collected in a 61 ms burst every $\sim$ 5 min. Narrowband (TBN) provides a stream of 100 kSPS allowing for a single tuning of 100 kHz (75 kHz \footnote{The filter rolloff removes 25 kHz of bandwidth.} usable) to be collected continuously. DP is configured such that the TBN data is duplicated between PASI and a data recorder (DR). This allows data to be both recorded for later processing and correlated in near real-time by PASI.


\section{Software}

\subsection{Correlation}
Correlations of the digitized voltages are achieved using a software FX correlator (PasiFx) written using a highly modular design to facilitate testing and modification of multiple components in parallel by multiple developers. 

The initial FX correlator code was developed at Los Alamos National Laboratory as part of their contribution to the LWA and was modified further at the University of New Mexico (UNM). The code's primary tasks are divided between the four nodes in the cluster, and it is highly threaded to take advantage of the multiple CPU cores per node.


Data can be input into PasiFx (Fig. 1) from three different sources: (1) Self-generated simulated antenna data, which produces Gaussian noise plus the signals from a number of user-specified point sources in the sky, (2) a stream of TBN data from DP, through the DR, via the 10 GbE connection to the DR, and (3) raw TBN data saved to a disk, which was mainly used during the testing and commissioning phase and is useful for debugging. Normal PASI operation uses the live stream of TBN. 

PASI receives the stream of TBN time series at a rate of 357 GB/hr. However, due to a limited number of inputs into DP one of the antennas from the main array is not used, resulting in the correlation of 260 dual polarization antennas. The voltages are then Fourier transformed and all 33,670 pairs are correlated to produce the full polarization products XX, YY, XY, and YX. Each visibility covers 100 kHz of bandwidth (75 kHz usable) with eight 12.5 kHz channels and are integrated for 5 seconds\footnote{Integrations for near real-time imaging can be as fine as 1 s, however the CPU burden at $<$5 s causes overheating. Therefore, PASI has almost exclusively been run at 5 s integrations.}. For each integration PasiFx outputs a CASA measurement set, containing all the visibilities. Figure 2 shows the snapshot $(u,v)$ coverage of the inner core of the LWA1, and Figure 3 shows perpendicular $(l,m)$ slices of the model PSF generated from the $(u,v)$ snapshot in Figure 2.

The final data rate of the visibilities is 5 GB/hr, roughly 1\% of the data rate of TBN into PASI.  The visibilities are saved for $\sim$4 months on a 15 TB ring buffer. This allows for a considerably lengthy time frame in case an interesting transient is found in the image data and reanalysis of the visibilities is needed.


\begin{figure}
	\centering
	\includegraphics[width = 7in]{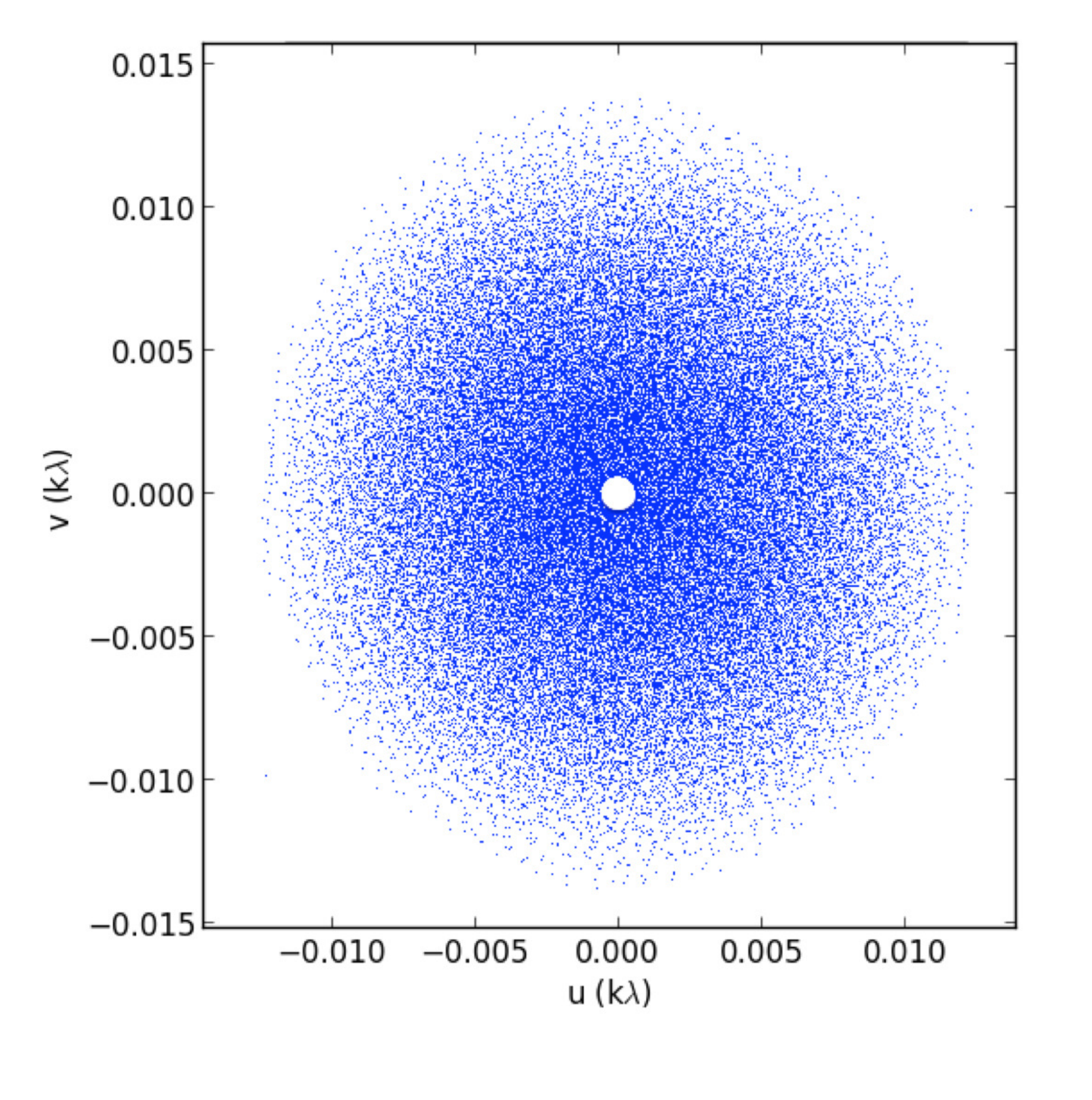}
	\caption{A plot of the $(u,v)$ coverage of the main core of the LWA1 at 38 MHz, u and v are in units of kilo wavelength. The inner hole is a result of the required minimum 
separation between dipoles of 5 meters \citep{Kogan09}. }
\end{figure}

\begin{figure}
	\centering
	\includegraphics[width = 7in]{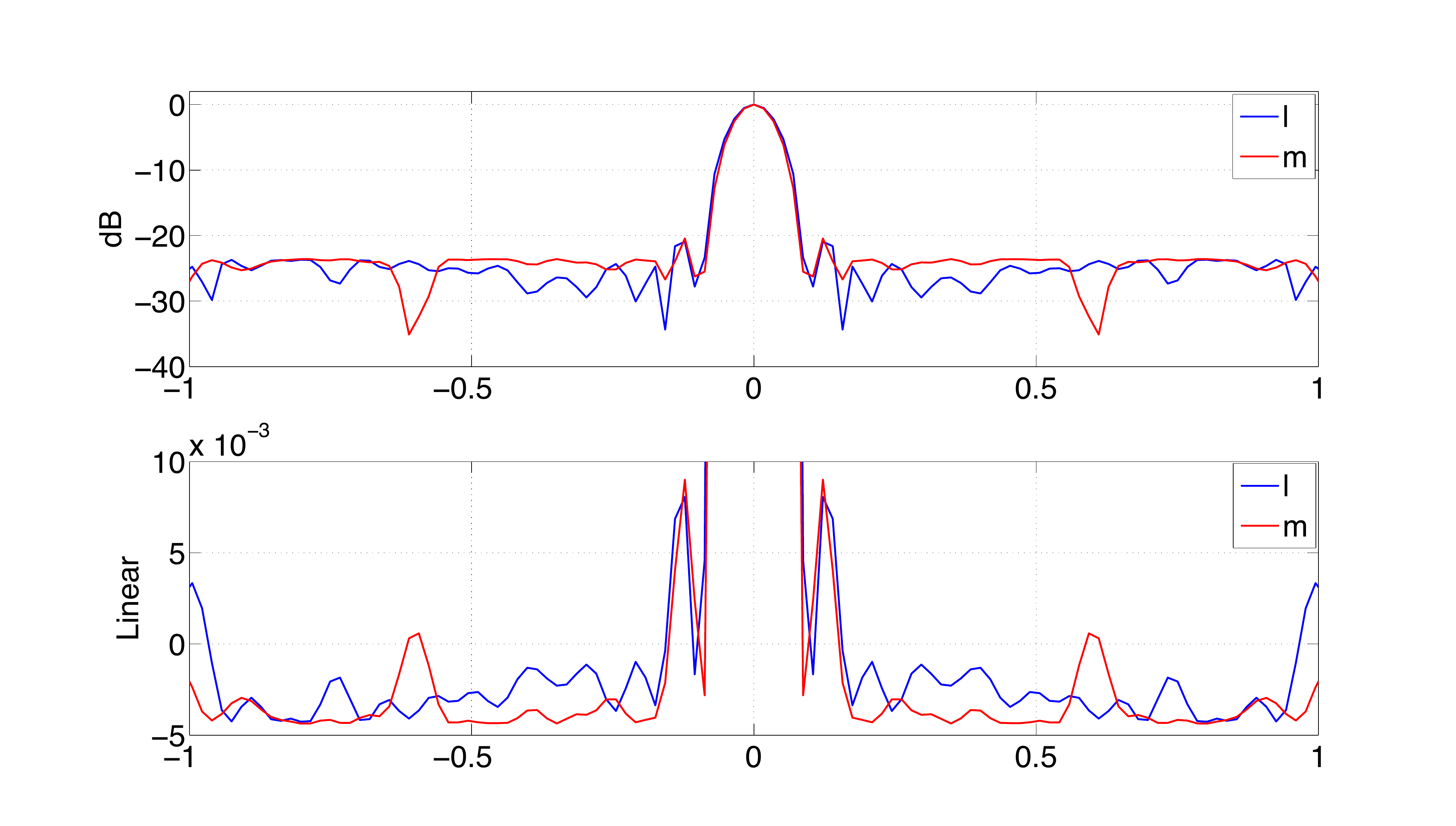}
	\caption{Perpendicular $(l,m)$ slices of the normalized, modeled point spread function taken from the $(u,v)$ snapshot at 38 MHz in Figure 2. The slices are shown in both absolute value dB (top) and linear (bottom) scaling. The units of $l$ and $m$ are directional cosines covering the full 180$^{\circ}$ from horizon to horizon, -1 to 1. The red line covers all $l$ where $m$ = 0. Likewise the blue line covers all $m$ where $l$ = 0.}
\end{figure}

%


Before correlation PASI also extracts and saves the averaged spectrum from all the antennas for each 5 s interval, with 256 390 Hz channels. The averaged spectra are used for identifying bright radio frequency interference (RFI) when searching for transients.   

\subsection{Imaging}

Due to a lack of computational resources and input sky models PASI only produces dirty images of the uncalibrated visibilities\footnote{All antenna signal paths have been delay calibrated and these delays are used by the correlator.  However, no additional phase or amplitude calibration, i.e., self calibration, is applied to the visibilities.}. The measurement sets are imaged using the Fourier transform and summing features of the CASA CLEAN script \citep{McMullin07}, but without any deconvolution using the CLEAN algorithm (niter = 0). We phase the visibilities to zenith so that the sky is projected onto a 2D circle with zenith at the center. We also use natural weighting with no $(u,v)$ taper. The pipeline produces dirty images for each of the four Stokes parameters and is executed within a Python-based pipeline outside of the normal CASA-Python interface. Each image is averaged over the inner six channels, which cover the 75 kHz of usable bandwidth. The image and cell size depends on the frequency, for 38 MHz we use 128$\times$128 pixels with a cell size of $1^{\circ}\times1^{\circ}$, we use larger image size and smaller cell size for higher frequencies and vice versa for lower frequencies.


The images, averaged spectra, and metadata are saved within a binary file onto an external disk. Metadata includes a time stamp, integration length, central frequency, and bandwidth.  New files are generated every hour or every time the center frequency is changed. The images are also uploaded to the LWA-TV website where the Stokes I and Stokes V images, as well as the averaged spectrum, are displayed in near real-time. Figure 4 shows an example of the LWA-TV display with an equatorial coordinate grid overlaid.

\begin{figure}
	\centering
	\includegraphics[width = 7in]{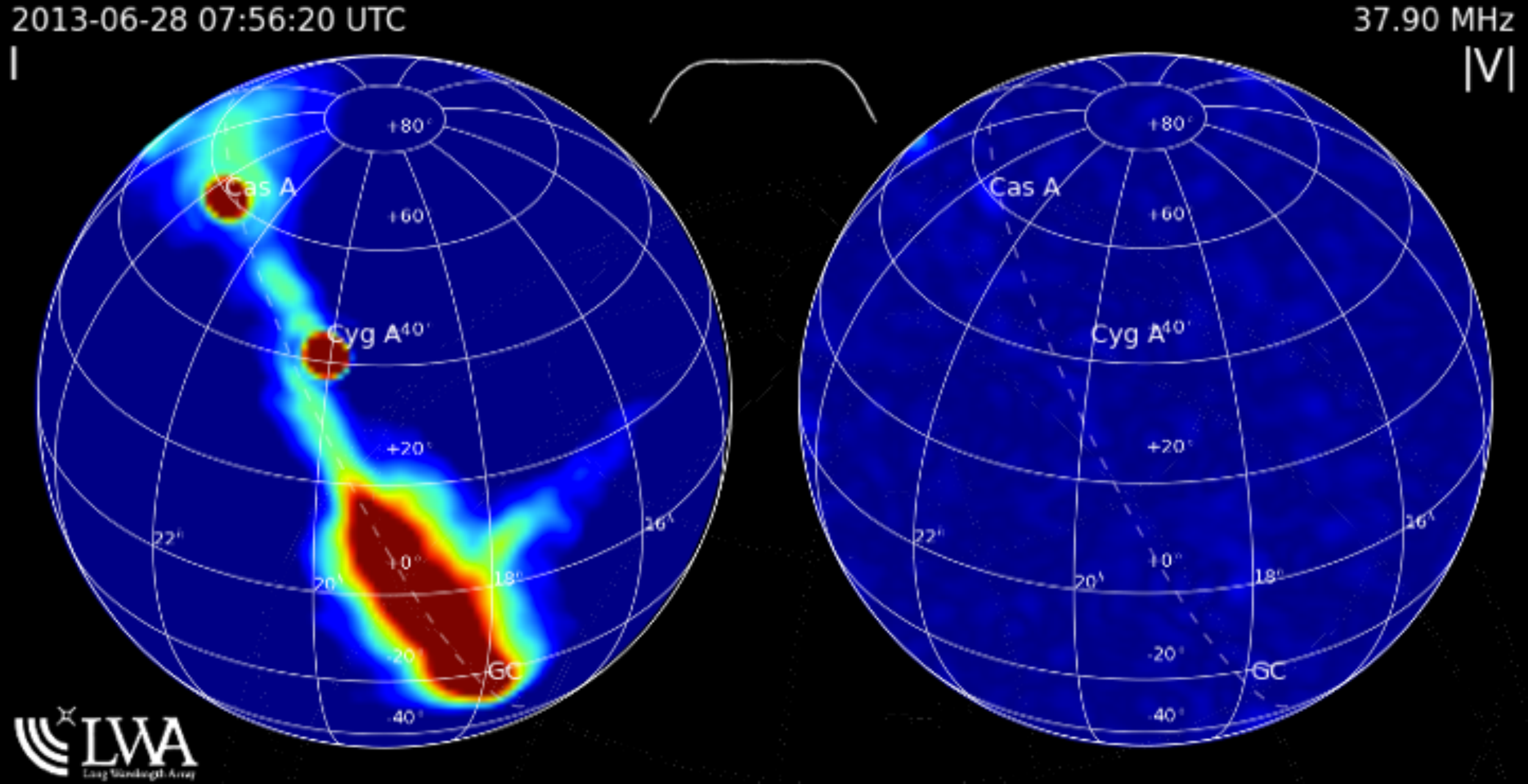}
	\caption{Example of a PASI 5 second snapshot centered at 37.9 MHz as displayed on the LWA-TV website, overlaid with a gird showing equatorial coordinates. The Stokes I image is shown on the left with a dynamic range of $\sim$ 300. The Stokes V image is shown on the right, with values shown as the absolute value so that both left hand circular (LHC) and right hand circular (RHC) are shown as positive. These images show the full $2 \pi$ sr of sky above the LWA1. The Galactic plane, Galactic Center (GC), Cygnus A (Cyg A), and Cassiopeia A (Cas A) can all be seen in the Stokes I image. The 256 channel averaged spectrum is shown in the top center of the figure. }
\end{figure}

The final data rate for the images is 200 MB/hr, roughly 0.06\% of the raw TBN. This data rate is small enough to allow for a permanent archive with infrequent disk swaps. While the initial intent of PASI was to run 24 hours a day, this cannot be accomplished. Down time is primarily due to the fact that currently beam observations degrade TBN observing when both modes operate in parallel. Since April 2012 PASI has been observing an average of 16 hours a day, and the image archive currently contains over 13,000 hours of data. 

\section{Data and Analysis}

The previous sections contain all the processing implemented automatically by PASI. Post processing requires user implementation either from a local computer or remotely from PASI. 

\subsection{Flux Calibration}
Flux calibration for PASI is determined for different frequencies as a function of elevation. The primary beam pattern of PASI is essentially the combined beam patterns of the two dipole antennas of an individual antenna stand. To measure this we simply extract the observed power received for Cygnus A and fit it with a 3rd order polynomial. Figure 5 shows the measured normalized power pattern at 38 MHz of Cygnus A and Cassiopeia A. The data was chosen from seven days across a year to show annual and diurnal consistency. This figure also shows the polynomial model derived for Cygnus A. The same procedure was carried out at 52 and 74 MHz, and the derived models were used when calibrating other sources as well as calculating the root mean square (RMS) of the image noise. The models were scaled using interpolated flux densities from the VLSS Bright Source Spectral Calibrator\footnote{http://www.nrl.navy.mil/rsd/vlss/calspec/} \citep{Helmboldt08}, which is based on values from \citet{Baars77}. The values used for Cygnus A at 38, 52, and 74 MHz are 25.5, 21.215, and 17.25 kJy.

It should be noted that the measured flux of any given source is affected by both the diffuse galactic emission surrounding it and the side lobes of other sources in the image. In particular, large negative bowls arise due to the combined side lobes of bright sources and the diffuse galactic emission. As can be seen in Figure 3 much of the PSF of the dirty beam is negative; therefore the added contribution from many sources can create negative bowls. The depth, size, and location of these negative bowls depends on what sources are present in the sky. Since Cygnus A and Cassiopeia A, the two point sources which contribute the most to the negative bowl, do not contribute to the negative bowl in their own locations, they are inherently less affected. Moreover since they are an order of magnitude brighter than the next brightest point source, they are fractionally affected least by the negative bowls. Other sources such as Taurus A and Virgo A are sufficiently affected by negative bowls that they do not accurately portray the primary beam pattern, and are therefore not used when calibrating.



%
%

\begin{figure}
	\centering
	\includegraphics[width = 7in]{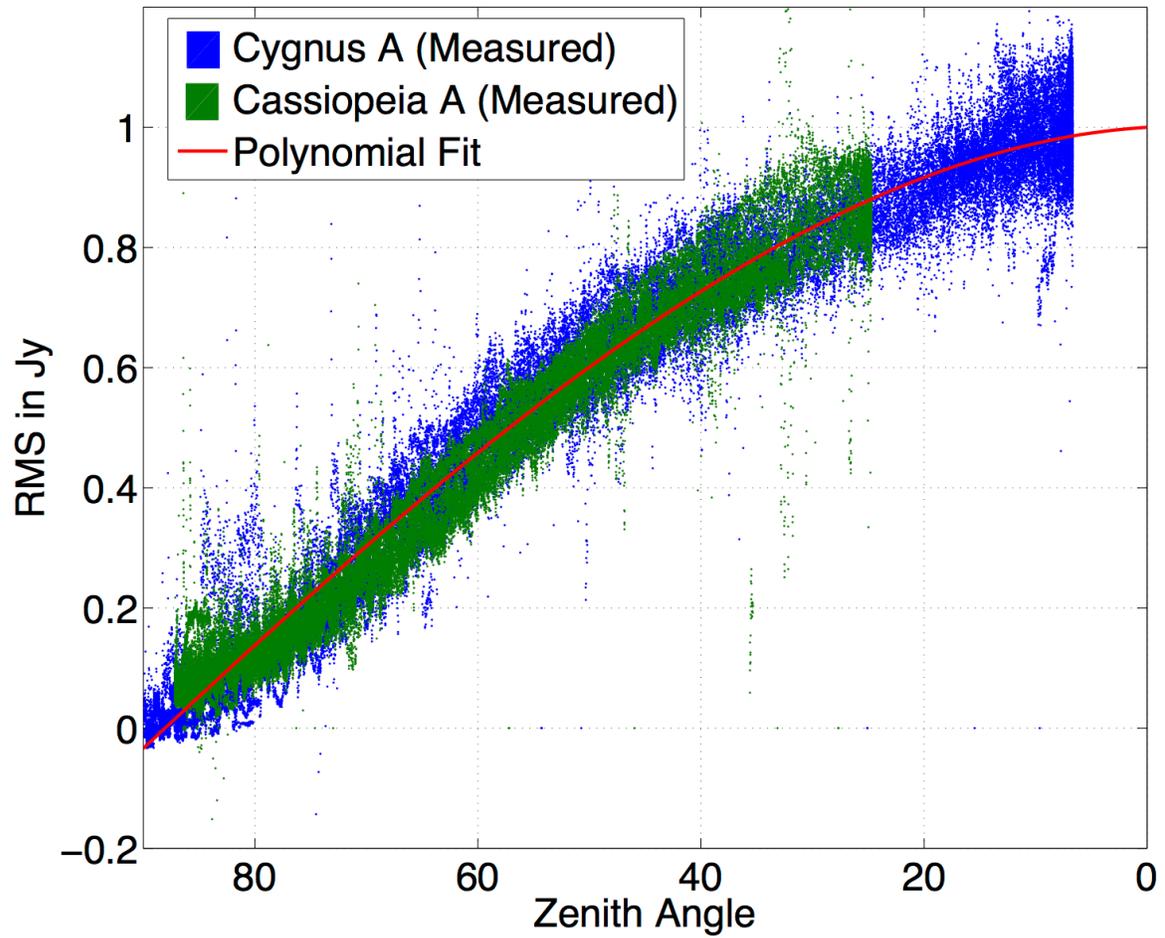}
	\caption{Normalized flux values for Cygnus A (Blue) and Cassiopeia A (Green) at 38 MHz, taken as the sources transited on seven separate days between June 2013 and June 2014. The polynomial model is shown in red. }
\end{figure}


Over 90\% of the data collected by PASI have been at center frequencies of 38, 52, and 74 MHz, with 65\%, 14\%, and 12\% recorded at those frequencies respectively. Therefore, flux calibration and sensitivity estimates have been obtained at these frequencies only.

\subsection{Sensitivity}

The zenith angle dependent RMS noise for PASI is calculated by calibrating the noise from pixels spread every degree from east to west running through zenith. This method provides estimates of 47 $\pm$ 13, 34 $\pm$ 11, and 45 $\pm$ 11 Jy  at zenith for 38, 52, and 74 MHz. The large range ($\pm$ 13 and $\pm$ 11)  in these values is due to confusion from the Galactic plane. Figure 6 shows the calculations of the RMS noise at  38 MHz at zenith as a function of Galactic latitude. It is clear that classical confusion from the Galactic plane contributes greatly to the noise at low Galactic latitudes. Moreover,  the fact that there is also some dependence far away from the Galactic plane also suggest that we are affected by side lobe confusion as well.

\begin{figure}
	\centering
	\includegraphics[width = 7in]{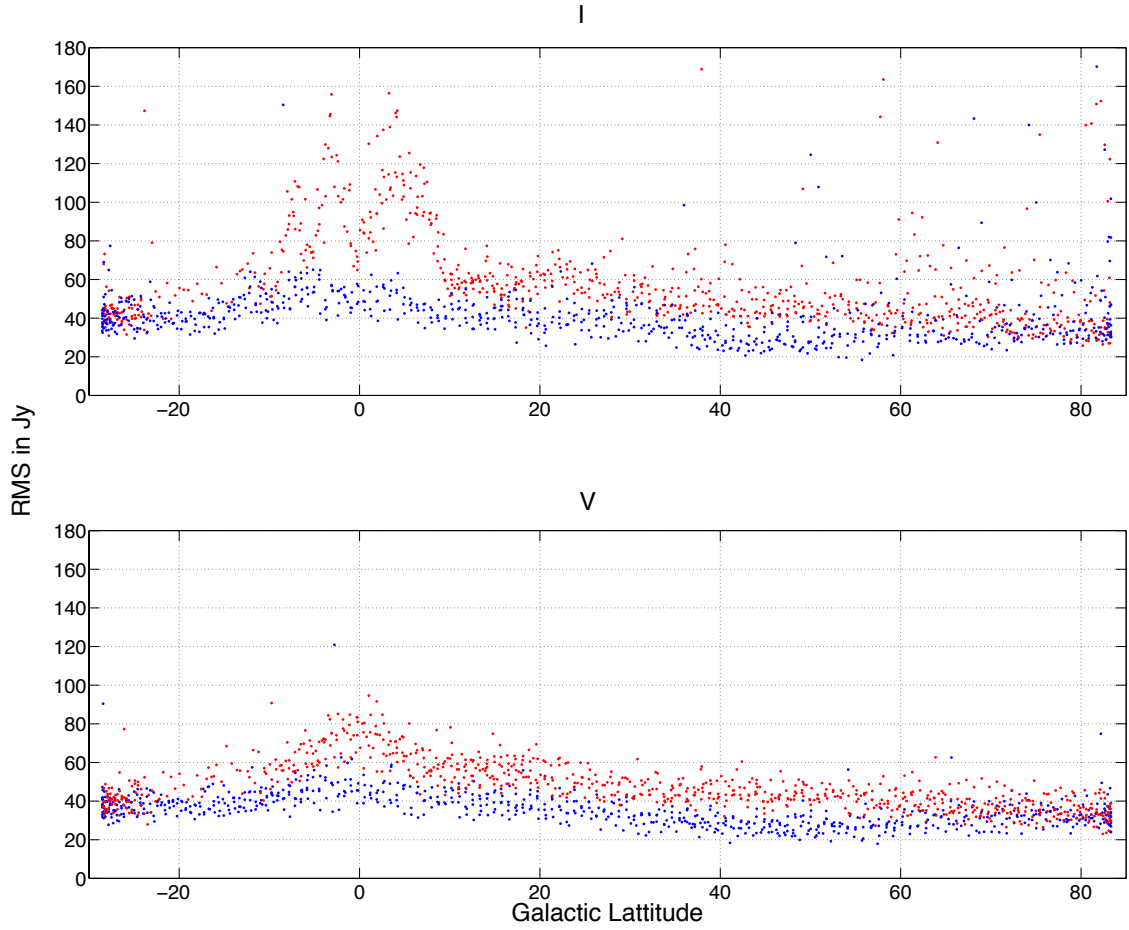}
	\caption{Calculations of the RMS at 38 MHz as a function of Galactic latitude for both Stokes I and V. The red dots correspond to Galactic longitudes $<120^{\circ}$ and the blue dots correspond to Galactic longitudes $>120^{\circ}$ . }
\end{figure}

Median values for the zenith angle dependent RMS at 38, 52 and 74 MHz are shown in Figure 7. The inverse of the polynomial models used for the power pattern are also shown. These models fit very well to the RMS data, but deviate at large zenith angles at 38 and 52 MHz most likely due to RFI that was not removed from the data. 

\begin{figure}
	\centering
	\includegraphics[width = 7in]{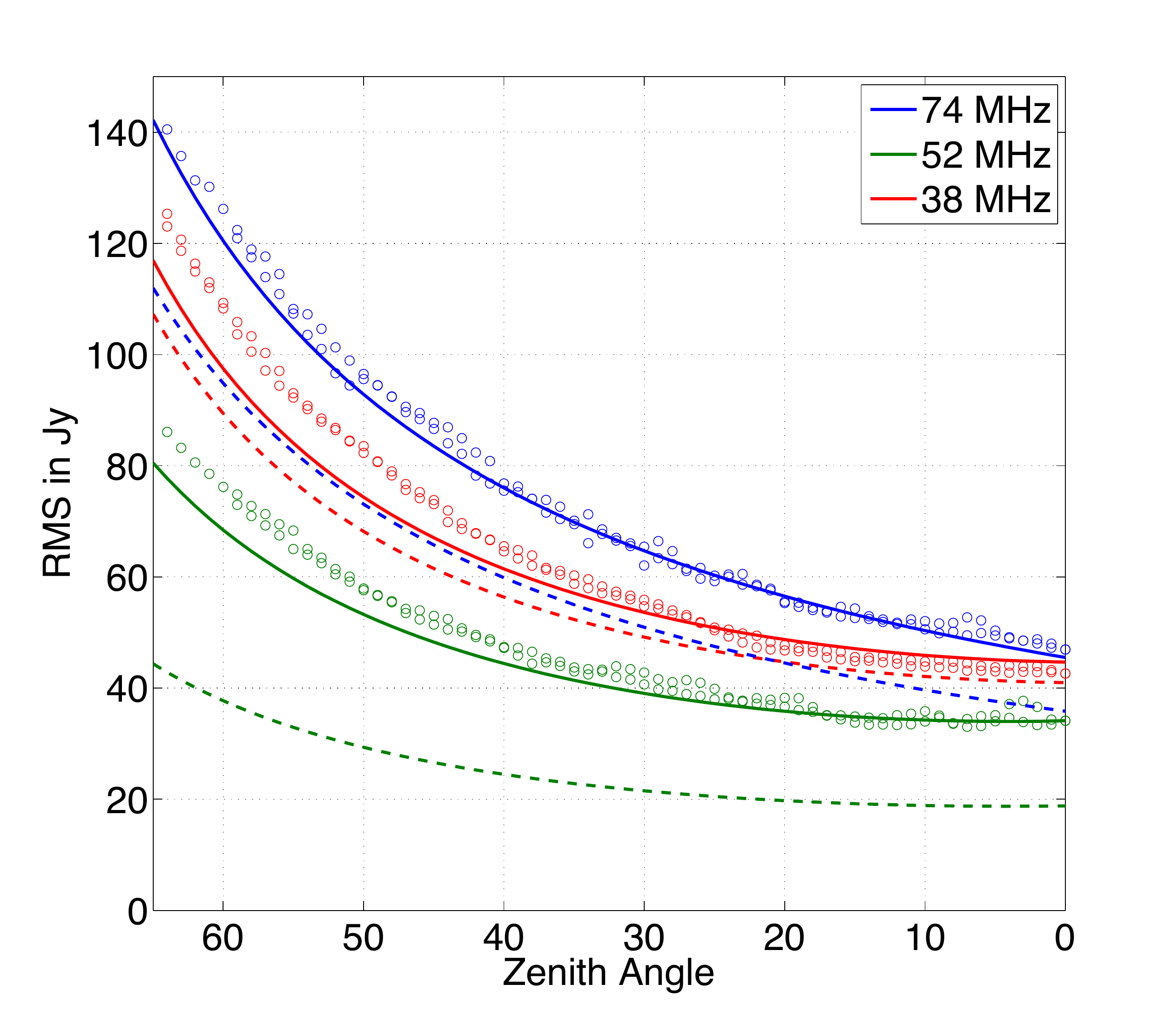}
	\caption{Shown here are the median values for the RMS noise for regular PASI images (open circles), the inverted polynomial model (solid lines), and the RMS of the subtracted images (dashed lines) for 38, 52, and 74 MHz. }
\end{figure}

Derived from beam measurements, estimates for the system equivalent flux density (SEFD) of the LWA1 range from 5 to 15 kJy, for frequencies between 36 and 80 MHz \citep{Schinzel14}. Given these values, the estimated RMS sensitivity for PASI at zenith with 5 second integrations and 75 kHz of bandwidth should range from 6 -  17 Jy, which is significantly lower than our measurements.

There are many possibilities as to this discrepancy. For instance there could be phase and amplitude errors in the uncalibrated visibilities in addition to classical and side lobe confusion from bright sources. Certainly any variation in the gain from the receiver chain could introduce errors to the measured RMS noise. This variation in gain has been measured and is mainly influenced by the temperature of the low noise amplifiers both in the shelter where the analog receivers are located, as well as at each of the antenna stands where the pre-amplifiers are located. Therefore the analog gain is affected by both the temperature inside the shelter as well as the ambient temperature to which the pre-amplifiers are exposed. On short time scales the air conditioning units are cycling on and off introducing a measurable variation in the gain of typically around 1\% \citep{Schinzel13}. The seasonal variation of ambient temperatures has a much larger effect on receiver gains introducing variations of up to 10\% over the course of a year, which was demonstrated by monitoring the received power of Cygnus A \citep{Schinzel15}. 

For this paper, however, the seasonal effects were not removed, instead we used a spread of days across the year. When we fit the 3rd order polynomial, the seasonal variations were averaged out. Similarly since we used the same data for measuring the RMS and used the median values, there should not be any systematic offset for the measured RMS. Indeed even if there was an offset, the maximum effect would be on the order of 10\%, which is much smaller than the near factor of 3 difference between our measurements and the beam measurements of \citet{Schinzel14}. Therefore gain variations cannot be the reason sole behind this discrepancy.

It is interesting to note that the Stokes V images have only slightly less noise than the Stokes I images. Since there is a lack of circularly polarized sources it is expected that the confusion contribution to the noise should go away. However, the Stokes V zenith sensitivities are 40 $\pm$ 11, 27 $\pm$ 8, and 41 $\pm$ 9 Jy, at 38, 52, and 74 MHz respectively. Moreover they too display a similar Galactic longitude dependance as the Stokes I data. See Figure 6. As will be discussed in the next section it is unlikely that instrumental leakage could be responsible for this since the leakage into Stokes V is $< 5\%$. One possibility is that ionospheric scintillation is adding noise to the data. As will be shown in \S 5.4, the scintillation seen in PASI is often circularly polarized. Since \citet{Schinzel14} used averaged power ratios to determine the SEFD, they would not have been sensitive to noise caused by ionospheric scintillation, since it would average out in their observations. Further research into the RMS noise is necessary to fully understand the discrepancy between the imaged and beam formed data.


\subsection{Polarization Leakage}

PASI saves images from all four Stokes parameters, but in order to measure the true polarization of a source, characterization of the instrumental leakage into each of these modes needs to be measured. Polarization leakage into the LWA1 dipoles is a function of both azimuth and elevation. By computing the ratio of power in Stokes V to Stokes I for a bright, unpolarized source one can measure the amount of leakage in that direction. However, there are only a handful of sources bright enough to perform such measurements, and they do not cover an extremely large range of individual azimuth/elevation coordinates. Therefore, a full characterization is not possible with PASI. 

Nevertheless we have measured the polarization leakage seen in the directions of Cygnus A and Cassiopeia A as they transit the sky. Figure 8 shows the percent leakage as a function of zenith angle for each Stokes parameter at 38 MHz. The leakage into the linear modes becomes significant at large zenith angles, but the leakage into the circular modes is hardly measurable. For all polarizations the mean leakage never rises above 15\%, and for Stokes V the leakage never gets above 5\% for zenith angles $<60^{\circ}$.

\begin{figure}
	\centering
	\includegraphics[width = 7in]{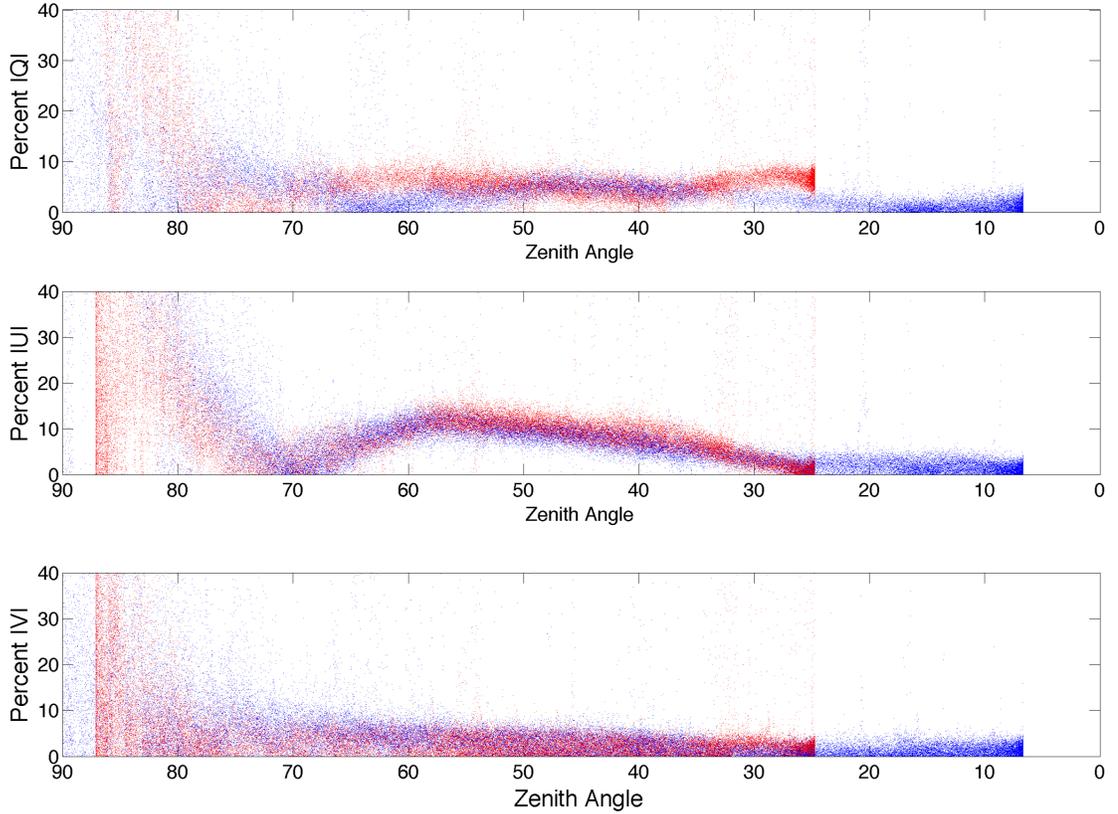}
	\caption{The measured magnitudes of polarization in Stokes Q, U, and V for unpolarized sources Cygnus A (Blue) and Cassiopeia A (Red), showing the amount of leakage into each Stokes parameter as a function of zenith angle. These data were taken from four different days at 38 MHz, one from each season of the year, showing no significant difference from one season to the next. While these two sources alone do not give a full picture of the azimuthal/zenith angle dependence of the polarization leakage, as plotted they do provide an estimate of the expected leakage as a function of zenith angle.}
\end{figure}

\section{Transients}
As mentioned in \S 1 transient searches with PASI have resulted in the detection of non-thermal radio emission from fireballs as well as limits on prompt radio emission from GRBs. PASI has also been used to place limits on astronomical transients, and this section describes the transient detection pipeline and the results thereof.  

\subsection{Image Subtraction}
Image subtraction algorithms have been developed for the purpose of finding transient sources on time scales of 5 s to several minutes \citep{Obenberger141,Obenberger142}. From every image a running average of the previous four images is subtracted, removing all steady sources. The images are then searched for pixels which exceed the 6 $\sigma$ level. The measured image noise values at zenith for subtracted images at 38, 52, and 74 MHz are 41, 19 and 35 Jy and their dependance on zenith angle is shown in Figure 7. For 38 MHz there does not appear to be much sensitivity improvement when using image subtraction. On the other hand, RMS noise at 52 and 74 MHz are vastly decreased when using image subtraction. The fact that the 38 MHz only slightly improves may point to ionospheric scintillation as the cause of decreased sensitivity. This effect would certainly be worse at lower frequencies and would most likely be observed on the order of several seconds. Therefore it may not get entirely removed by our image subtraction scheme.


\subsection{Radio Frequency Interference}

Depending on the frequency, time of day, and time of year PASI is subjected to varying amounts of RFI. A large fraction of the RFI seen by the LWA1 is narrow band \citep{Obenberger11}, which we can avoid by tuning TBN to protected or limited bands. Within the frequency range that PASI operates there are four protected bands designated for radio astronomy: 13.36 - 13.41, 25.55 - 25.67,  37.5 - 38.25, and 73.00 - 74.60 MHz. These protections allow for an exceptionally clean environment at 38 and 74 MHz, however 13 and 25 MHz can at times be significantly affected by interference, presumably from out of band emission from transmitters at other frequencies. 

When RFI does appear in PASI it is almost always located on the horizon and is easily removed by ignoring zenith angles greater than 60$^{\circ}$. Moreover, RFI is almost always linearly-polarized and is narrower than PASI's 75 kHz bandwidth. Therefore, transient candidates are checked for both linear polarization as well as spectral features within the six channels. The PASI image files contain all four Stokes parameters, however to retrieve the six frequency channels, the visibilities need to be reimaged.

Along with narrow band RFI, PASI is also affected by broadband RFI from power lines and electrical equipment. These sources are easily identified because they come from fixed regions on the horizon. Situated only a few hundred meters to the north east of the array are power lines which feed the LWA1 and facilities of the VLA. These power lines create micro-sparks which emit broadband bursts and are typically only a problem during the dry windy days of spring.

\subsection{Lightning}
Long Wavelength interferometry has been used to explore the physics of lightning since the late 1970s \citep{Warwick79}.  Recently, with the advent of inexpensive time-keeping via GPS and improvements in signal processing, a number of dedicated systems such as the Lightning Mapping Array (LMA; \citealt{Rison99} ) have been deployed.  This has clearly demonstrated the regions in clouds where lightning flashes originate, as well as illuminating in three dimensions the charge structures of clouds \citep{Marshall05}. It has also been correlated with severe weather of all types and with radar measurements to produce a much deeper understanding of the relation between lightning and its parent storm.  These arrays generally make use of a small number of dipoles, which is reasonable given the strength of the emission, but leads to limitations in the imaging quality obtained. The LWA1 and the PASI back-end have collected over 40 hours of lightning observations.  A special lightning mode was developed to produce rapid (200 fps) imaging with PASI in post-processing of recorded TBN data.  In principle, this imaging could be extended down to the sampling rate of 100 ksps or 0.01 msec.    At 5 msec time-resolution (Fig. 9) one can see the step-like motion of negative breakdown inside a positively charged cloud region.  The entire sequence of this flash can be viewed at http://www.phys.unm.edu/$\sim$lwa/lwatv/lightning/056161\_000042789\_00254.400\_5ms.mov.

\begin{figure}
	\centering
	\includegraphics[width = 7in]{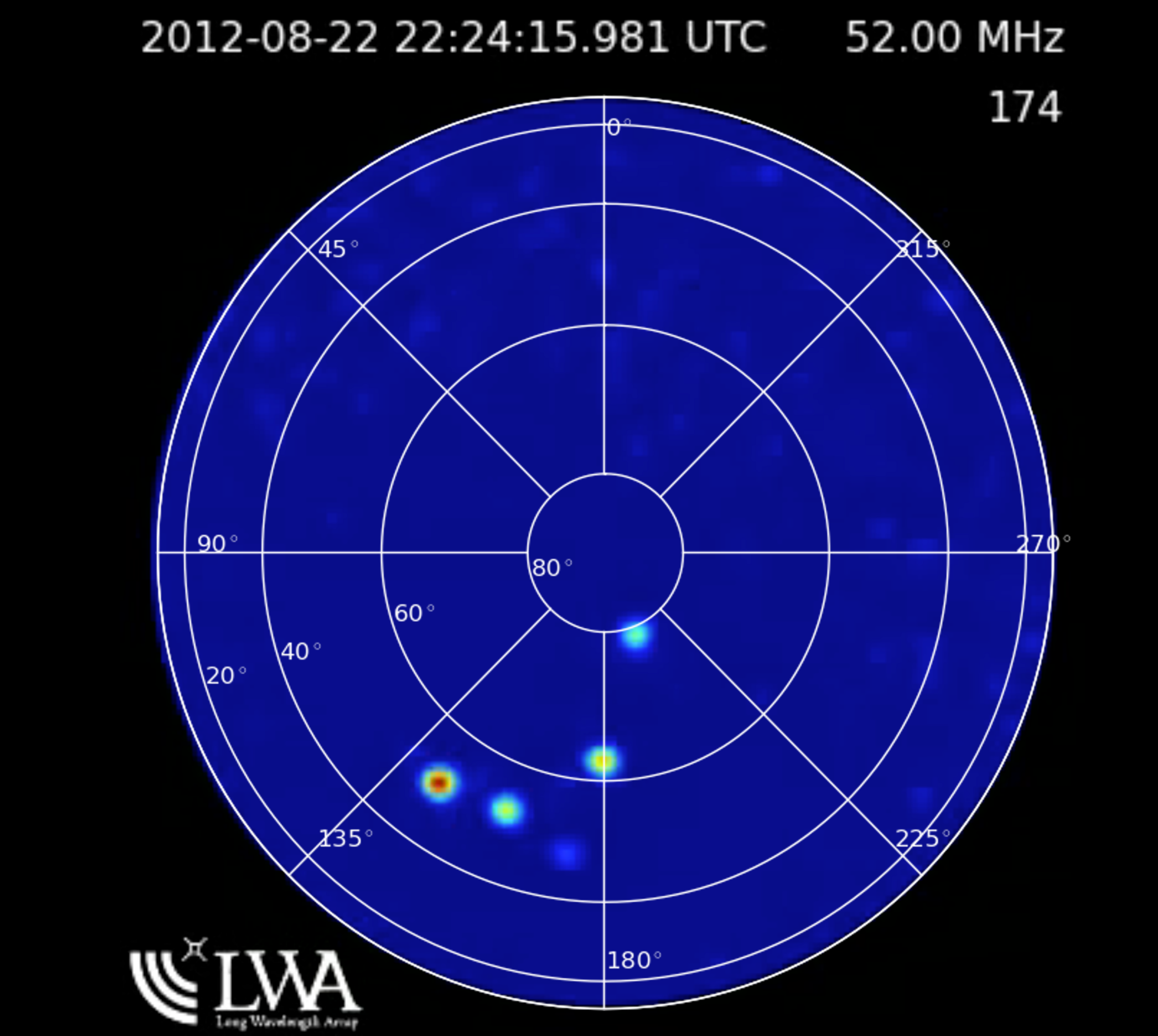}
	\caption{A 5 ms snapshot from PASI's lightning mode showing the step-like motion of the negative breakdown within a positively charged cloud region. Overlaid is the a grid showing Horizontal coordinates. }
\end{figure}



\subsection{Ionospheric Scintillation}
The transient search pipeline finds many transient sources which correspond to bright catalog sources, normally not detectable by the LWA1 because their flux densities are well below the noise floor. These events are not inherent to the sources themselves, but are thought to be caused by scintillation by the Earth's ionosphere \citep{Obenberger141}. The scintillation often appears in both the Stokes I and V images. See Figure 10. This is most likely due to the fact that the ionosphere is a magnetized plasma and therefore displays birefringence for the LHC and RHC polarizations, and therefore magnifying them separately. In addition to the amplitude scintillations we have observed position shifts in sources of up to $\sim$ 10$^{\circ}$ in extreme conditions. These shifts along with the amplitude scintillation may be responsible for the remaining subtracted image vs beam formed sensitivity discrepancy described in \S 5.1.

\begin{figure}
	\centering
	\includegraphics[width = 7in]{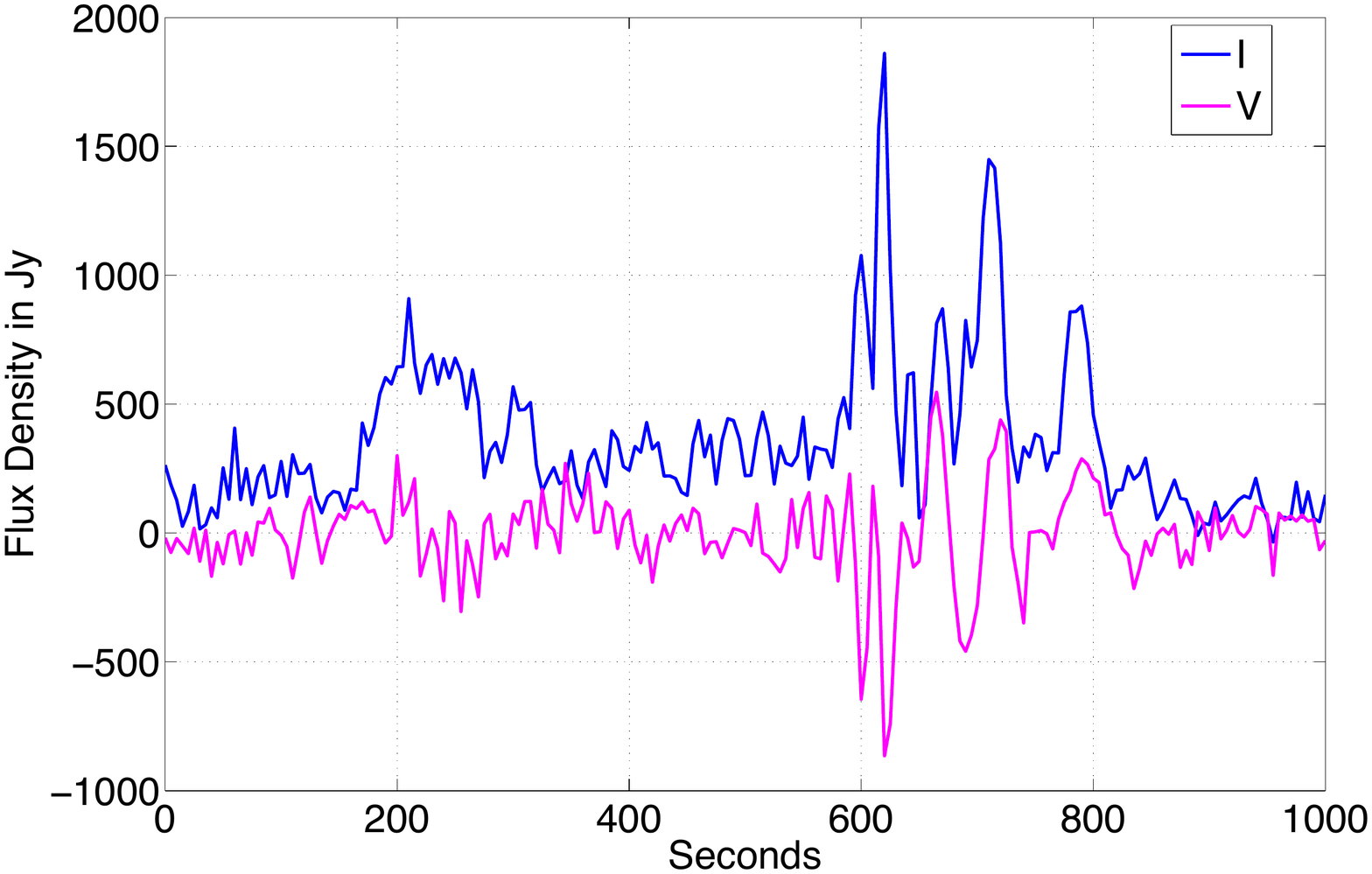}
	\caption{A light curve showing the scintillation of 3C254, a 60 Jy source at 38 MHz \citep{Kellermann69}. The scintillation can be seen in Stokes V, where positive values correspond to RHC and negative values correspond to LHC. This cannot be due to instrumental leakage, which is $< 5\%$ and does not rapidly switch between LHC and RHC. }
\end{figure}

Because of this scintillation, the regions within 1.5$^{\circ}$ of known bright sources are ignored in the transient search pipeline. A catalog of these sources is provided by the VLA low frequency sky survey (VLSS; \citealt{Cohen07,Lane12}). Excluding the $\sim$ 350 sources that are $>20$ Jy at 74 MHz removes $\sim$ 10\% of the sky observable from the LWA1, but sufficiently suppresses these false detections. 


Likewise the diffuse emission from the Galactic plane and north polar spur is focused by the ionosphere. Rather than excluding these regions, we increase the threshold for transient candidates to 10 $\sigma$ for Galactic latitudes of -7$^{\circ}$ to +7$^{\circ}$ and 8 $\sigma$ for latitudes from -10$^{\circ}$ to -7$^{\circ}$ and +7$^{\circ}$ to +10$^{\circ}$. For the north polar spur we set a threshold of 7 $\sigma$ for Galactic latitudes from  +10$^{\circ}$ to +70$^{\circ}$ and Galactic longitudes from +20$^{\circ}$ to +45$^{\circ}$. While these regions are not completely excluded the potential for discovering transients here is lessened. These regions comprise 25\% of the sky observable from the LWA1.

\subsection{Astronomical Transients }

To date PASI has found 60 transients between 25.6 and 37.9 MHz, within 13,000 hours of data recorded between 25.6 and 74.0 MHz. Evidence suggests that a large fraction, if not all, of these transients are bursts of non-thermal radio emission from fireballs \citep{Obenberger142}. Assuming that none of these sources have a different origin, this data places limits on the rate density of astronomical transients that are an order of magnitude lower than previously set by similar studies at these frequencies \citep{Cutchin11,Kardashev77,Lazio10}.

Since 91\% of the data recorded by PASI has been at 38, 52, and 74 MHz, we only provide transient limits for these frequencies. While PASI has an extensive amount of sky coverage, the sensitivity is not uniform across the sky. As shown in \S 5.1 the sensitivities of subtracted PASI images at zenith are 41, 19, and 35 Jy at 38, 52, and 74 MHz and follow the zenith angle dependance shown in Figure 6. At a zenith angle of 40$^{\circ}$ the RMS sensitivities are 56, 24, and 60 Jy, while at a zenith angle of 60$^{\circ}$ the RMS sensitivities are 90, 38, and 95 Jy. However, the portion of the sky with zenith angles $<40^{\circ}$ includes 23\% of the sky above the LWA1, while zenith angles  $<60^{\circ}$ includes 50\%. 

Therefore, using image subtraction we place rate density limits on transients at 38, 52, and 74 MHz occurring at zenith angles $<60^{\circ}$ brighter than the thresholds of 540, 230, and 570 Jy (6$\sigma$ at z = 60$^{\circ}$). With 8,400 hours recorded at 38 MHz and 8,353.7 deg$^{2}$ of sky observable on thresholds\footnote{At zenith angles $<60^{\circ}$, the flux density thresholds at each frequency begin to include the regions within the Galactic plane that are held at higher S/N thresholds because the sensitivity gets better at smaller zenith angles. Combining the regions removed for bright sources and the high threshold set for diffuse emission excludes 19\% of the sky.} of $>$ 540 Jy, the transient rate density is limited to $<1.2\times10^{-4}$ yr$^{-1}$ deg$^{-2}$. These limits are for 5 s integrations, and we assume that the transients are 5 s rectangular pulses and that the entire duration of each pulse lies within one integration. Under these assumptions the pulse energy density limit to $>2.7\times 10^{-23}$ J m$^{-2}$ Hz$^{-1}$. 

Similarly for 1,900 and 1,400 hours recorded at 52.0 and 74.0 MHz, the rate density for 5 s transients at those frequencies is limited to $<5.6\times10^{-4}$ and $<7.2\times10^{-4}$ yr$^{-1}$ deg$^{-2}$ with pulse energy densities of $>1.1\times 10^{-23}$ and $>2.8\times 10^{-23}$ J m$^{-2}$ Hz$^{-1}$. Table 1 shows these values with comparison to similar studies at these frequencies. PASI has decreased the existing rate density limits by an order of magnitude for comparable pulse energy densities at these frequencies. 

It should be noted that no dedispersion is applied on the PASI data, therefore we begin to lose sensitivity to transients with high dispersion measures (DM) due to dispersion smearing \citep{Obenberger141}. For 5 s bursts at 38, 52, and 74 MHz, the S/N drops\footnote{The factor of 1.5 was chosen as a reference because it is roughly half the range of the zenith angle dependent RMS function.} by $>$ 1.5 at $DM$s $>$  220, 570, and 1,600 pc cm$^{-3}$. However, if the bursts are longer than 5 s then dispersion smearing has a smaller effect on the sensitivity. For instance, if the burst lasted 30 s then the S/N would drop by 1.5 at DMs of 1,300, 3,400, and 9,800 pc cm$^{-3}$.

\begin{deluxetable}{ |c| |c| |c| |c| |c| }
\tablecaption{Pulse Rate Limits}
\tablewidth{0pt}

\tablehead{\colhead{Name} 		&	\colhead{Frequency}	&	\colhead{Rate Density}		&	\colhead{Pulse Energy Density}		&	\colhead{Pulse Width}\\    	&	  \colhead{(MHz)}	&	  \colhead{(yr$^{-1}$ deg$^{-2}$)}	&	  \colhead{(J m$^{-2}$ Hz$^{-1}$)}  }

\startdata
Kardashev et al. (1977)	&	60			&	$10^{-3}$ 			&	$3.1\times 10^{-22}$			&	0.5 s\\
                                      	&	38			&	$1.5\times10^{-3}$	&	$2.1\times 10^{-22}$ 		&	0.5 s\\
\hline
Lazio et al. (2010)		&	73.8			&	$10^{-2}$ 			&	$1.5\times 10^{-20}$ 		&	300 s\\
\hline
Cutchin (2011)			&	38			&	$2.5\times10^{-1}$	&	$2.6\times 10^{-23}$			&	3 s\\
\hline
This Paper			&	38			&	$1.2\times10^{-4}$	&	$2.7\times 10^{-23}$			&	5 s\\
                                             	&	52			&	$5.6\times10^{-4}$	&	$1.1\times 10^{-23}$			&	5 s\\
					&	74			&	$7.2\times10^{-4}$	&	$2.8\times 10^{-23}$ 		&	5 s\\					
\enddata

\end{deluxetable}

\clearpage

%
%
%
%
%
%

\section{LWA TV}

LWA TV has been viewed in numerous countries, and in a recently we have been receiving $\sim$ 2000 unique IP hits per month. While most of these hits are presumably members of the public at large, we know of several cases in which LWA TV has been used during astronomy lectures.  Dedicated\footnote{To learn how to set up a dedicated monitor for viewing LWA TV, please visit http://fornax.phys.unm.edu/lwa/trac/wiki/EPO} monitors displaying LWA TV have been installed at the VLA Visitor Center and in the lobby of the Physics and Astronomy department at UNM.

 LWA TV has also been used to localize and mitigate interference at the site.  In one instance some of the fluorescent light fixtures in the antenna assembly building of the VLA were found to be emitting by noticing that they switched on at the start of the work day and switched off at the end.  Several times now we have used LWA TV to localize and mitigate power line noise (recognizable by its presence at the horizon and its extent) working with the Socorro Electric Cooperative.

\section{Conclusions}
This paper has described the system design of PASI, a backend correlator of the LWA1 telescope. Also reported here are the results from a transient search covering 10,313 deg$^{2}$ (1 $\pi$ sr) of the sky for 13,000 hours at a relatively unexplored region of the electromagnetic spectrum. This is the deepest search for transients below 100 MHz and has resulted in new limits on radio transients at 38, 52, and 74 MHz. At these frequencies, 5 s pulses with pulse energy densities  $>2.7\times 10^{-23}$, $>1.1\times 10^{-23}$, and $>2.8\times 10^{-23}$ J m$^{-2}$ Hz$^{-1}$ are limited to rate densities of $<1.2\times10^{-4}$, $<5.6\times10^{-4}$, and $<7.2\times10^{-4}$ yr$^{-1}$ deg$^{-2}$.

PASI has also been used to characterized radio transient foregrounds such as emission from fireballs and ionospheric scintillation, which will need to be taken into account in future transient studies with the LWA1 as well as other low frequency telescopes. It also offers the potential for future studies of the ionosphere, lightning, and meteors.

\section{Acknowledgments} We thank the anonymous referee for thoughtful comments. 

Construction of the LWA1 has been supported by the Office of Naval Research under Contract N00014-07-C-0147. Support for operations and continuing development of the LWA1 is provided by the National Science Foundation under grants AST-1139963 and AST-1139974 of the University Radio Observatory program. 

This research has made use of the NASA/IPAC Extragalactic Database (NED) which is operated by the Jet Propulsion Laboratory, California Institute of Technology, under contract with the National Aeronautics and Space Administration.

Part of this research was carried out at the Jet Propulsion Laboratory, California Institute of Technology, under a contract with the National Aeronautics and Space Administration.

The PASI computing cluster was provided by a grant from the New Mexico Consortium.

This research received funding from Los Alamos National Laboratory LDRD project number 20080729DR and the Institute for Advanced Studies.

This work has been authored by an employee of Los Alamos National Security, LLC, operator of the Los Alamos National Laboratory under Contract No. DE-AC52-06NA25396 with the U.S. Department of Energy. The United States Government retains and the publisher, by accepting this work for publication, acknowledges that the United States Government retains a nonexclusive, paid-up, irrevocable, world-wide license to publish or reproduce this work, or allow others to do so for United States Government purposes.  Los Alamos National Laboratory strongly supports academic freedom and a researcher's right to publish; however, the Laboratory as an institution does not endorse the viewpoint of a publication or guarantee its technical correctness. This paper is published under LA-UR-14-26948.

\bibliographystyle{plainnat}

\end{document}